\documentclass[aip,prd,showpacs,floatfix,preprint,a4paper]{revtex4-1}
\usepackage{amsmath,bm}
\usepackage{graphicx}
\usepackage{commath}
\usepackage{epsfig}
\usepackage{cancel}
\usepackage{color,soul}
\begin{document}

\title{Properties of rapidly rotating hot neutron stars with antikaon condensates at constant entropy per baryon}

\author{Neelam Dhanda Batra$^1,^2$, Krishna Prakash Nunna$^1$ and Sarmistha Banik} 
\affiliation{$^1$Dept. of Physics, BITS Pilani, Hyderabad Campus, Shameerpet Mandal, Hyderabad-500078, India.\\  $^2$Dept. of Physics, Indian Institute of Technology, New Delhi-110016, India.}



\date{\today}

\begin{abstract}
We consider a neutrino-free hot neutron star that contains antikaon condensates in its core and is at finite entropy per baryon. We find the equation of state for a range of entropies and antikaon optical potentials and generate the mass profile of static as well as rotating stars. Rotation induces many changes in the stellar equilibrium, and hence its structural properties evolve. In this work, we report the effect of rotation on the mass and shape of a hot neutron star for different equations of state and thermodynamic conditions. Temperature profile of hot, static neutron star is also explored. We also make a crude estimate of the amplitude of gravitational waves emitted by an axisymmetric rotating NS with high magnetic field. 
\end{abstract}

\maketitle




\section{Introduction}

The matter inside the core of a neutron star (NS) is super dense. 
Here, the baryon number density ($n_b$) can reach up to a few times of normal nuclear matter density ($n_0$), which is unlike anything found on Earth. The behaviour of matter up to nuclear densities has been studied and is well documented by numerous nuclear physics experiments. Lack of experimental data at higher matter densities means we do not completely understand the properties of matter at super nuclear densities. Consequently, there exist huge uncertainties in understanding the behaviour of matter at high densities such as those present in NS cores.

NS are the nature's laboratories for studying such highly dense matter. One indirect way to study the properties of such matter is by knowing the mass profile of a NS. The NS mass is fairly established by many observational studies.  The maximum masses observed till now are $1.928\pm 0.017$ $M_{solar}$ and $2.01 \pm 0.04\: M_{solar}$ for PSRs J1614-2230 and J0348+0432 respectively {\cite{demo, anton, fons}. Any  cold equation of state (EoS) intending to describe the matter inside the core of a NS must be able to reach upto this level. This constraint in itself rules out many of the proposed EoS for NS. Knowing mass alone is, however, insufficient to rigidly pinpoint  towards underlying EoS and a knowledge of NS radius is very much required. Further, there exist inherent uncertainties in the determination of a NS radius observationally \cite{Pot14,Stein}. A quest to understand the NS EoS is a wide open field of research at present and provides an important link in comprehending the behaviour of matter at high densities.

The major constituents of matter with density just below $n_0$, are protons, neutrons and leptons. The constituents of higher density matter remain uncertain to a large extent due to the lack of experiments. As a result, the interactions between the constituents of such matter are poorly understood. Many studies have theorised the appearance of hyperons at higher matter densities \cite{Gle,debarati,Oertel16,Marq17,Lonar15}. It has also been suggested that Bose-Einstein condensates such as those of pions and/or antikaons are favourable to appear at highly dense NS cores \cite{Ginz65,Kap86,pons}. The  appearance of new particles such as $K^-$ condensates results in the softening of EoS as negatively charged leptons are replaced by the slow-moving massive condensates which do not contribute to pressure. The overall pressure thus increases less steeply with density, thereby resulting in a softer EoS. This, in effect, lowers the maximum mass reached by a NS \cite{pons,meta,third}.  

A NS is born in a core-collapse supernovae (CCSN) explosion, which is believed to be adiabatic, i.e., the entropy per baryon ($s$) of each mass element remains constant during the collapse except during the passage of the shock \cite{Lieb05}. About fifty seconds after birth, the stellar interior becomes fully transparent to the neutrinos \cite{PrakashPR97}. Some processes, e.g., frictional dissipation of the rotational energy or Ohmic decay of the internal magnetic field, may reheat stellar interior thus delaying the cooling, especially at these late stages \cite{Yak}. The exotic composition of the
cores such as quarks or kaon/pion condensates, affects the neutrino emission mechanism and hence its cooling properties \cite{Yak}. Higher the threshold density, slower will be the cooling of compact stars via kaon condensation \cite{Schaab}. The temporal evolution of  static proto-neutron stars (PNS) has been thoroughly studied by Pons et. al. (1999)  \cite{proto}. In this work, we are interested in studying a rotating NS that is born in an adiabatic environment. We assume a range of isentropic profile for the hot star. The maximum value of $s$ reached by a PNS can be $1-2k_B$ \cite{proto}, which may increase to $5k_B$ for high mass progenitor\cite{Hempel2012} or merger of NS \cite{Kast16}. For simplicity, we have restricted ourselves to a deleptonised star at constant $s$ \cite{Gouss}, so that neutrinos do not contribute to the lepton number of the matter. We consider a hot star containing exotic matter such as $K^-$ condensates and that is yet to cool down to Fermi temperature.

NS are mostly observed in radio and are known as pulsars (PSRs). Pulsar observation has come a long way since the first PSR was detected by Jocelyn Bell Burnell and Antony Hewish in the year 1967 \cite{Hewish67}. Till date, more than $2500$ PSRs have been observed in our galaxy \cite{ATNF}, out of which a large fraction has time period between 0.1 to 10 seconds and are known as ordinary PSRs. The millisecond pulsars (MSP) on the other hand are part of binary systems and are rapidly spinning stars that are powered by the flow of matter and momentum from their companion stars. The 24 known AXMSP (accreting X-ray milli second PSRs) have an average time period of $\sim 3$ ms. When X-ray emission ceases, these stars emit radio waves. There are $\sim 300$ recycled radio MSP in our galaxy\cite{Tauris}. The fastest known pulsar, PSR J1748-2446ad \cite{Hess06}, has a rotation period of 1.397 ms and a frequency of 716 Hz. In addition, some young PSRs are also observed to be rotating with fast velocities. For example, the Crab pulsar rotates once in 33 ms and is known to emit radio giant pulses. The frequency of a rotating star is limited by the mass-shedding phenomenon only. At mass shedding limit a rigidly-rotating NS rotates with maximum frequency possible which is defined as its Kepler frequency.

 Soon after the CCSN explosion, the  neutrino-trapped PNS is expected to be rotating differentially due to lack of enough viscous forces \cite{Gous98, Ott06}. Differentially rotating stars can support significantly more mass in equilibrium than static or rigidly rotating stars \cite{Shap00}. Keeping in mind the uncertainty in initial rotational state of collapsing core, the actual degree of non-uniformity in rotation of a PNS is unknown. As they settle into $\beta$-equilibrium, viscosity dampens the differential rotation. Apart from the slight differential rotation following glitches,  NS are expected to rotate uniformly \cite{Fried}.  Here, we have considered an idealised scenario of uniform and rigid rotation about an axisymmetric axis which represents an approximation to the actual rotational state of a hot NS \cite{Gouss}. In our calculations, we consider only neutrino-less hot NS matter.
 
In general, the equilibrium of a rotating NS depends on the rotational effects considerably. The mass-radius relation for a static NS has been established theoretically by the well known Tolman-Oppenheimer-Volkoff (TOV) equations which give an upper bound to the mass of a static NS.  The internal structure of a NS changes as it rotates with higher and higher frequencies. As the centrifugal force increases with increase in rotational velocity, a rotating star can support larger mass compared to a static one. The rotating stars also tend to have larger radii \cite{ijmpd}. This change is not only due to the appearance of centrifugal force but also due to the 'frame-dragging' of inertial reference frames \cite{Gle, Thir18}.  However, at finite temperature the pressure never really vanishes and hence the surface of the NS cannot be determined definitely. The maximum mass reached in a sequence is also a function of the constituent composition as well as the temperature profile inside a NS. In this paper, we restrict our discussions to the mass and temperature of a NS.

We study the rotating NS sequences with EoS containing exotic particles and having different entropies. These results are compared with the corresponding static configurations and also with the nucleon only matter configuration for a better understanding. Further, we study the relativistic equilibrium configurations of rotating hot NS with different EoS and thermodynamic conditions, in terms of their fluid energy density profiles. We also study how they are affected by the change in rotation frequency (or angular momentum) upto the Keplerian limit. In the present work, we only consider NS after it has deleptonised and wherein the rigid rotation has set in due to viscosity, but is yet to cool down to Fermi temperatures.

We also make a rough estimate of gravitational wave (GW) amplitude emitted by a NS having strong magnetic field $(B)$. Most NS are the superdense remnants of supernova explosions but some are formed in binary NS merger as well. Some NS may have very high magnetic fields and are called magnetars. The origin of high $B$ in such NS is an open ended problem. Simulations show that a magnetar which is formed after the merger of binary NS, is differentially rotating and ultraspinning with typical periods of the order of a few milliseconds and magnetic field strengths in the range of\cite{Giaco2013} $B \sim 10^{15}- 10^{16} G$. The NS magnetic field ($B\sim 10^{12} G$) is amplified by several orders of magnitude ($B\geq 2 \times 10^{15} G$) within the first millisecond after merger \cite{Price2006}. The differential rotation can further increase the field. Further, the long term evolution models have shown that the magnetic field can lead to a uniformly rotating NS surrounded by an accretion disk and with a collimated magnetic field \cite{Duez2006}.} On the other hand, observations have shown that the NS that are relatively old, have strong magnetic fields which are of the order of $B \sim 10^{11}$ G to $10^{13.5}$ G, but much longer periods (P $\sim 1$ s). Whereas, millisecond radio pulsars have ultra-fast rotation (P $\leq 20$ ms) and much weaker magnetic fields ($B \leq 10^{10}$ G)\cite{JOAA16}. Since we are interested in young and hot NS, we take a typical magnetar formed in a merger event for the estimation of GW amplitude.

This paper is organised in the following way. In sections \ref{sec1a} and \ref{sec1b} we describe our model EoS for a  compact star. In section \ref{section: LORENE} we discuss the rotation and axisymmetric deformation in such a star.  Section \ref{sec:result} contains our results and the related explanation. Finally, in section \ref{sec:concl} we conclude with a summary of results and further research work being done in continuation. In this paper we have used natural units with $k_B=1$, wherever required.


\section{Models of Compact stars}\label{sec1}
\subsection{Equation of state of core}\label{sec1a}
\label{sec:EOS}
We consider nuclear and $K^-$ condensed matter in the dense interior of a NS and calculate the EoS within the framework of relativistic mean field (RMF) model with density-dependent  coefficients. The nucleons (N); denoted by spinors $\psi_N$; have mass $m_N$ and interact through exchange particles $\sigma$, $\omega$, and $\rho$ mesons. The density-dependent RMF model Lagrangian density for the nucleons is given by \cite{dd2_14},
\begin{eqnarray}\label {eq_lag_N}
{\mathcal L}_N &=& \sum_N \bar\psi_{N}\left(i\gamma_\mu{\partial^\mu} - m_N
+ g_{\sigma N} \sigma - g_{\omega N} \gamma_\mu \omega^\mu 
-  g_{\rho N} 
\gamma_\mu{\mbox{\boldmath $\tau$}}_N \cdot 
{\mbox{\boldmath $\rho$}}^\mu  \right)\psi_N\nonumber\\
&& + \frac{1}{2}\left( \partial_\mu \sigma\partial^\mu \sigma
- m_\sigma^2 \sigma^2\right)
-\frac{1}{4} \omega_{\mu\nu}\omega^{\mu\nu}
+\frac{1}{2}m_\omega^2 \omega_\mu \omega^\mu \nonumber \\
&& - \frac{1}{4}{\mbox {\boldmath $\rho$}}_{\mu\nu} \cdot
{\mbox {\boldmath $\rho$}}^{\mu\nu}
+ \frac{1}{2}m_\rho^2 {\mbox {\boldmath $\rho$}}_\mu \cdot
{\mbox {\boldmath $\rho$}}^\mu.
\end{eqnarray} 
Here, 
$
\omega^{\mu \nu}= \partial^ \mu \omega^ \nu-\partial^\nu \omega^ \mu$
and
${\boldsymbol \rho^{\mu \nu}}= \partial^ \mu \rho^ \nu-\partial^\nu \rho^ \mu $
are the field strength tensors for the vector mesons, and  
${\mbox{\boldmath $\tau_{N}$}}$ is the isospin operator.

The meson-baryon couplings $g_{\alpha N}(\hat n)$'s ($\alpha=\sigma,~\omega$ and $\rho$) depend on vector density $\hat n = \sqrt{j_{\mu}j^{\mu}}$, where
${ j}_\mu = \bar \Psi_N \gamma_\mu \Psi_N$. The $g_{\alpha N}(\hat n)$'s are 
Lorentz scalar functionals of baryon field operators and are determined following the prescription of Typel (2005)\cite{typel05} and Typel et. al. (2009) \cite{typel09}. They reproduce the bulk properties of nuclear matter such as nuclear compressibility, symmetry energy and its slope parameter at saturation density, corresponding to the density dependence of symmetry energy \cite{dd2_14, Lim}. The details of the parameters used in this calculation can be found in Char and Banik (2014) \cite{dd2_14}. This parameterisation is known as DD2 formalism. In the mean field approximation, the nucleon-meson couplings become a function of total baryon density $n_b$, i.e., 
$<g_{\alpha N}(\hat n)>=g_{\alpha N}(<\hat n>)=g_{\alpha N}(n)$ 
\cite{typel09,ddrh}. 
The density-dependence of meson-baryon couplings \cite{ddrh} gives rise to the rearrangement term 
$\Sigma^{(r)}_N$, the expression for which can be written as, 

\begin{equation}\label{eq_rear}
\Sigma^{(r)}_N=\sum_N[-g_{\sigma N}'  
\sigma n^{s}_N + g_{\omega N}' \omega_0 
n_N
+ g_{\rho N}'\tau_{3N} \rho_{03} n_N ]~.
\end{equation}
Here $g_{\alpha N}'=\frac {\partial g_{\alpha N}} {\partial n_N}$, 
$\alpha= \sigma,~ \omega, ~\rho$ and $\tau_{3N}$ is the isospin projection of $N=n,p$.
We compute the dense matter EoS of the NS in the mean-field approximation, where the meson fields are replaced by their expectation values.  The time-like components of vector fields and the isospin components of $\rho$ fields survive in a uniform matter. The mean-fields are denoted by  $\sigma$, $\omega_0$ and $\rho_{03}$, where 
$m_{\sigma}^2\sigma = g_{\sigma N} (n_n^S +n_p^S)$, 
$m_{\omega}^2\omega_0 = g_{\omega N} (n_n +n_p)$ and 
${m_\rho}^{2}\rho_{03}={1\over2}g_{\rho N}\left(n_{p}-n_{n}\right)$; 
 $n_p, n_n$ and $n_p^S, n_n^S$are the number densities and scalar number densities of proton and neutron respectively. The number density of nucleon at finite temperature is given by,

\begin{equation}
n_N = 2 \int \frac{d^3k}{(2\pi)^3} \left({\frac{1}{e^{\beta(E^*-\nu_N)} 
+ 1}} - {\frac{1}{e^{\beta(E^*+\nu_N)} + 1}}\right). 
\end{equation}
Here, $\beta = 1/T$ and 
$E^* = \sqrt{(k^2 + m_N^{*2})}$.
Scalar density for nucleons on the other hand is,

\begin{equation}
n_N^S = 
2 \int \frac{d^3 k}{(2\pi)^3} \frac{m_N^*}{E^*} 
\left({\frac{1}{e^{\beta(E^*-\nu_N)} 
+ 1}} + {\frac{1}{e^{\beta(E^*+\nu_N)} + 1}}\right) ~.
\end{equation}

The Dirac equation for the interacting nucleons is given by
$[\gamma_{\mu}\left(i\partial^{\mu}-\Sigma^{\mu(r)}_{N}\right)-m_{N}^*
]\psi_{N}=0~.$
The effective nucleon mass is defined as 
$m_N^*=m_N-g_{\sigma N}\sigma$. 
The chemical potential for the nucleon is
$\mu_{N} = \nu_N + g_{\omega N} \omega_0 + g_{\rho N} \tau_{3N}\rho_{03} + \Sigma^{(r)}_N,~ $ where $\nu_N= \sqrt{k^2 + m_N^{*2}}$.
The pressure due to nucleons is calculated as following  \cite{Gle},

\begin{eqnarray}
P_N &=& -\frac{1}{2} m_\sigma^2 \sigma^2
+ \frac{1}{2} m_\omega^2 \omega_0^2 
+ \frac{1}{2} m_\rho^2 \rho_{03}^2  + \Sigma^{(r)}_N \sum_N n_{N}
\nonumber \\
&& + 2T \sum_{N=n,p} \int \frac{d^3 k}{(2\pi)^3} 
[ln(1 + e^{-\beta(E^* - \nu_N)}) +
ln(1 + e^{-\beta(E^* + \nu_N)})]. 
\end{eqnarray}
The explicit form of the energy density is given below,
\begin{eqnarray}
\epsilon_N &=& \frac{1}{2}m_\sigma^2 \sigma^2
+ \frac{1}{2} m_\omega^2 \omega_0^2 
+ \frac{1}{2} m_\rho^2 \rho_{03}^2  \nonumber \\
&& + 2 \sum_{N=n,p} \int \frac{d^3 k}{(2\pi)^3} E^* 
\left({\frac{1}{e^{\beta(E^*-\nu_N)} 
+ 1}} + {\frac{1}{e^{\beta(E^*+\nu_N)} + 1}}\right)~.  
\end{eqnarray}
The rearrangement term does not contribute to the energy density explicitly.  However, it occurs in the expression for pressure through the baryon chemical potentials. It is the rearrangement term that accounts for the energy-momentum conservation and thermodynamic consistency of the system \cite{Fuchs, TypelW, ddrh}.

 We adopt the finite temperature treatment of antikaon condensates as given in Pons. et. al. (2000)\cite{pons}. They had considered three forms of Lagrangian.  It was argued that the divergence of constant vector fields is necessarily zero, making the form given by Knorren-Prakash-Ellis (KPE) \cite{Knorren} sufficient for the description of kaon sector \cite{PrakashPR97} and we stick to that choice. The antikaons are described by the Lagrangian density
${\mathcal L}_K = D^*_\mu{\bar K} D^\mu K - m_K^{* 2} {\bar K} K ~$,
where the covariant derivative is
$D_\mu = \partial_\mu + ig_{\omega K}{\omega_\mu}
+ i g_{\rho K} {\boldsymbol \tau}_K \cdot {\boldsymbol \rho}_\mu $ 
\cite{pons, meta, third}.  The isospin doublet for kaons is denoted by 
$K\equiv (K^+, K^0)$ and that for antikaons by $\bar K \equiv (K^-, \bar K^0)$.
It is to be noted that the antikaon-baryon couplings are density independent.
The effective mass of the antikaons is given by $m_K^* = m_K - g_{\sigma K} \sigma $,
where $m_K$ is the bare kaon mass. 
The in-medium energy of the $K^\pm$ mesons is given by, 
\begin{equation}
\omega_{K^\pm} = \sqrt{(k^2+ m_K^*)}  \pm (g_{\omega K} \omega_0 
+  g_{\rho K} \rho_{03}).
\end{equation}
For s-wave (${\bf k}=0$) condensation, the momentum dependence vanishes in $\omega_{K^\pm}$. The threshold condition for s-wave $K^-$ condition is $\mu_{K^-} =\omega_{K^-}$. The chemical equilibrium in the reaction $n \leftrightarrow p + K^-$ sets the chemical potential of $K^-$ as,

\begin{equation}
\mu_{K^-}= \mu_n - \mu_p =\mu_e.
\end{equation}
The $\omega_{K^+}$ never drops to meet the threshold condition of $\omega_{K^+} = \mu_e$. On the other hand, $\omega_{K^-}$ decreases from its vacuum value $m_K$ with increasing density as the meson fields grow and thus only $K^-$ condensates appear in the system \cite{PrakashPR97}. 

The net antikaon number density is given by, $n_K = n_K^C + n_K^{T}~$,
where, $n_K^C$ gives the $K^-$ condensate density. Here, $n_K^{T}$ represents the thermal
density and is given by,

\begin{equation}
n^C_{K} = 2\left( \omega_{K^-} + g_{\omega K} \omega_0
+ \frac{1}{2} g_{\rho K} \rho_{03} \right) {\bar K} K
= 2m^*_K {\bar K} K  ~.
\end{equation}

\begin{equation}
n_K^{T} =  
\int \frac{d^3 k}{(2\pi)^3} 
\left({\frac{1}{e^{\beta(\omega_{K^-}-\mu)} 
- 1}} - {\frac{1}{e^{\beta(\omega_{K^+}+\mu)} - 1}}\right)~.  
\end{equation}
The condensates do not contribute to pressure,  but implicitly change the rearrangement term of 
Eq. \ref{eq_rear} via the values of meson fields. The energy density of $K^-$ condensates is given by, 

\begin{eqnarray}\label{eq_epsilonK}
\epsilon_K &=& m_K^* n_K^C + \left( g_{\omega K} \omega_0
+ \frac {1}{2} g_{\rho K} \rho_{03} \right) n_K^T \nonumber \\
&&+
\int \frac{d^3 k}{(2\pi)^3} 
\left({\frac{\omega_{K^-}}{e^{\beta(\omega_{K^-}-\mu_{K^-})} 
- 1}} + {\frac{\omega_{K^+}}{e^{\beta(\omega_{K^+}+\mu_{K^+})} - 1}}\right)~.  
\end{eqnarray}

The first term in Eq. \ref{eq_epsilonK} is the contribution due to $K^-$ condensate and second and third terms are the thermal contributions to the energy density in $\epsilon_K$.

In addition to nucleons and $K^-$ mesons, we also have leptons in the system. 
They are treated as non-interacting particles and the relevant physical 
variables for EoS, i.e., number densities, energy densities and 
pressure are calculated following a similar method as that used for nucleons, using the Lagrangian density 
${\mathcal L}_l= \sum_l \bar\psi_l\left(i\gamma_\mu {\partial^\mu} - m_l \right)\psi_l ~.$
Here,  $\psi_l$ ($l \equiv {e,\mu}$) denotes the lepton spinor. In a NS, when the electron chemical potential $\mu_e$ becomes equal to the muon mass, the
electrons are converted to muons by $e^- \rightarrow \mu^- + {\bar \nu}_{\mu} + \nu_{e}$.  Therefore, in a NS the onset 
of muons is determined by the condition $\mu_e = \mu_{\mu}$. The muons are  usually ignored in hot dense matter
owing to their high rest mass ($m_{\mu} \sim 105.66 MeV/c^2$), which suppresses 
their formation. However, at high temperatures the electron chemical potential 
exceeds $m_{\mu}$ leading to a significant number of muons. Nevertheless, in our study,  the  energetically favoured antikaon condensates replace the leptons as soon as they are formed. We have ignored the $\mu^+$s , as their formation is highly suppressed.
The total energy density in the presence of $K^-$ condensates is therefore,
$\epsilon = \epsilon_N + \epsilon_K + \epsilon_l$. 

We generate the EoS at constant entropy per baryon ($s$). The entropy density (${\mathcal S}_N$) of nucleons and leptons is related to energy density and pressure through Gibbs-Duhem relation ${\mathcal S}_N = \beta \left(\epsilon_N + P_N - \sum_{i} \mu_i n_i \right)$, where, $i = n, p, l$. The entropy density of antikaons is, 
${\mathcal S}_K = \beta \left(\epsilon_K + P_K -  \mu_{K^-} n_K \right)$, where, $n_K =n_K^C + n_K^T$. The entropy per baryon is given by $s={\mathcal S}/n_b$, where $n_b$ is the total baryon density. The total entropy per baryon has contribution from the nucleons, antikaons and leptons i.e., $s=({\mathcal S}_N +{\mathcal S}_K +{\mathcal S}_l)/n_b$. 

\subsection{Matching different parts of the EoS}\label{sec1b}

 Hempel and Schaffner-Bielich constructed the HS(DD2) EoS for dense matter consisting of neutrons, protons and leptons. The low-density, inhomogeneous part of this EoS was calculated in the extended Nuclear Statistical Equilibrium model (NSE) \cite{hs1}, that we use for our purpose. It consists of non-uniform matter of light and 
heavy nuclei along with unbound nucleons at low temperatures and densities that are below nuclear saturation. Interaction among the unbound nucleons are described by considering the same Lagrangian density as in Eq. \ref{eq_lag_N} and using the density dependent formalism \cite{typel09,hs1}. As the $K^-$ condensates appear only at high densities and  at relatively high temperatures, the nuclei and exotic matter are never found to coexist.  Therefore, we simply use the non-uniform part of the  HS(DD2) EoS \cite{hs1, bhb} following the standard prescription of minimisation of free energy as is given in Banik et. al. (2014) \cite{bhb}. Although the above procedure allows for a smooth transition between the different parts of EoS at around nuclear saturation density, it is of course not completely consistent as was emphasised in Marques et. al. (2017)\cite{Marq17}. Recently, Fortin et. al. (2016) \cite{Fortin16} have shown that the core-crust matching does not have any effect on the maximum mass allowed for the star, 
but the uncertainty in radius calculations can be $\sim 4\%$ depending on the way core-crust matching is done. Since our emphasis is not on the calculation of radii of stars, this uncertainty does not affect our overall results. However, this EoS table \cite{bhb} is for supernova for a wide range of temperature, baryon number density and electron fraction. For NS matter, we impose an additional condition of $\beta$ equilibrium on the chemical potentials, $\mu_n-\mu_p=\mu_e$. For a given temperature and $n_b$, the electron fraction is determined by finding the zero of the function $f(Y_e)= \mu_e(Y_e)$ at a fixed value of s  \cite{compose}, where $Y_e$ is the electron fraction and $\mu_e$ is given by Eq. 8.

\subsection{Rotation and axisymmetric deformation of a NS}
\label{section: LORENE}
To compute and compare the hydrostatic equilibrium configurations of rotating NS with DD2 EoS as described above, we use {\it nrotstar} code of the numerical library {\it Lorene} \cite{LORENE} which implements multi-domain spectral method for calculating accurate models of rotating NS in full general relativity \cite{lor1}.

In this formalism, field equations are derived using 3+1 formulation. This forms a system of four elliptic partial differential equations, which are then solved numerically using the self-consistent-field method. While solving, the space-time is assumed to be asymptotically flat and axisymmetric. Under these simplified assumptions of space-time, the metric function can be written as,

\begin{equation}
g_{\alpha \beta}dx^{\alpha} dx^{\beta}= -N^2 dt^2 + A^2 \left( dr^2 + r^2 d \theta^2\right) + B^2 r^2 sin^2\theta {\left(d \phi - w dt\right) }^2.
	\label{eq_metric}
\end{equation}

Here, N, A, B and w are functions of (r, $\theta$). The accuracy of the solutions obtained using {\it nrotstar} is checked using the general relativistic virial theorem GRV3 and GRV2 \cite{lor2}, which gives a typical value of 
$\sim 10^{-4}$.

{\it Lorene/nrotstar} is formulated primarily for cold EoS, or a barotropic EoS
\cite{Gouss}. Our EoS are temperature dependent but are formulated so as to have constant entropy per baryon. This results in a homoentropic flow, thereby making the EoS barotropic. Thus as long as we have an EoS that is isentropic, {\it Lorene} formalism can be used to do the calculations.

Using {\it Lorene/nrotstar}, we compute stable NS configurations for different EoS as described in the previous section. We measure the change in mass profile of a NS as it rotates with different angular momenta. Finally, we make an estimate of the strength of GW, that can be emitted from a uniformly rotating NS whose magnetic field axis is not aligned with its rotation axis.

\section{Results}
\label{sec:result} 

We have generated a number of isentropic EoS profiles and calculated the properties of a reasonably rapidly rotating and deleptonised NS using DD2 model.  We consider nucleons-only system consisting of protons (p), neutrons (n) and leptons (l); and denote it by "np". When the matter consists of antikaon condensates ($K^-$) and thermal kaons ($K^T$), it is denoted by "npK". The potential depth of antikaons in saturated nuclear matter is given by, $U_{\bar K}= -g_{\sigma K} \sigma + g_{\omega K} \omega_0$. The study of kaonic atoms suggests an attractive optical potential for the antikaons. The value of $U_{\bar K}$ at $n_0$ has been calculated in a coupled-channel model and
chiral analysis of $K^-$ atomic and scattering data. However, till date, no definite consensus exists regarding the value of $U_{\bar K}$ \cite{ddrh, ramos}. For our calculations, we have chosen a wide range for $U_{\bar K}$, from a shallow value of -60 MeV to a deeper one of -150 MeV \cite{ddrh}.  The coupling constants for kaons at saturation density for different values of $U_{\bar K}$ in the DD2 model are listed in Table~1 of Char and Banik (2014) \cite{dd2_14}.

\begin{figure}
	\includegraphics[width=\columnwidth, height=8.5cm]{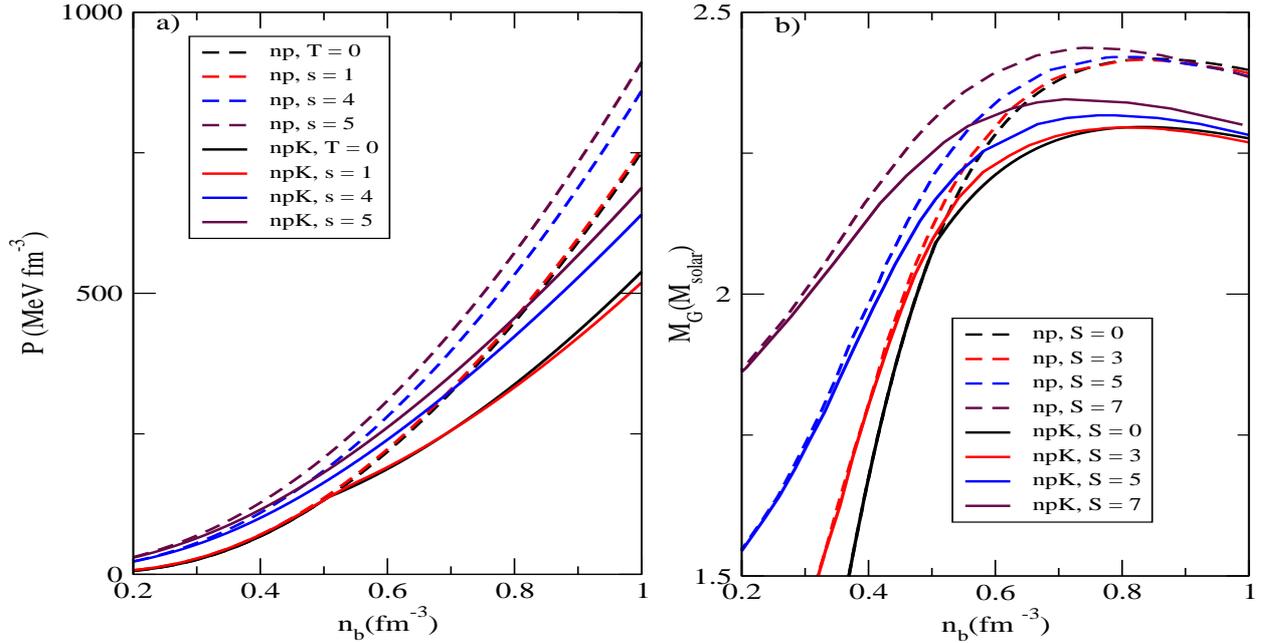}
    \caption{a) The EoS with pressure plotted against 
number density for np and npK ($U_{\bar K}$ = -100 MeV) for a NS core at zero temperature state (T = 0 MeV) and at adiabatic state (entropy per baryon $s$ = 1, 4 and 5
). 
b) The mass sequences against number densities for the np and npK ($U_{\bar K}$ = -100 MeV) EoS, for total entropy $S$ = 0, 3, 5 and 7 $M_{solar}$. The dashed lines are for np 
matter and solid lines are for npK matter for $U_{\bar K}= -100$ MeV in both the panels.}
    \label{fig1}
\end{figure}

In the left panel of Fig.\ref{fig1}, we plot pressure vs. baryon number density or the EoS profile for np and npK matter ($U_{\bar K}$ = -100 MeV) for different $s$ values. We notice that the set of npK EoS is softer compared to that of np matter. 
 As we go from np to npK matter, leptons in the NS core are gradually replaced first by thermal kaons and later by antikaon condensates also as the core density increases. These condensates do not contribute to the pressure term but they do contribute to the net negative charge in the system and hence the overall EoS becomes softer.
 
We also compare the hot NS EoS having different $s$ values (1, 4 and 5) with EoS for cold NS matter (T=0 MeV). In general, for a given composition, the EoS is softer for a NS with lower $s$. This matches with the previous results for other model EoS \cite{PrakashPR97}. At a given density, higher $s$ matter has higher chemical potential and hence higher pressure, as is evident from the Gibbs-Duhem relation.  
We further notice that the EoS profiles for $s=1$ NS matter is only slightly different than that for cold NS matter for np, but this difference between the two EoS is more evident for npK composition. The np EoS is slightly stiffer for matter at $s=1$ than for matter at zero temperature, as expected, because kinetic pressure increases due to increase in temperature, but the potential pressure term remains the same.
In contrast, for npK, $s = 1$ EoS is comparatively softer than  zero temperature EoS, especially at higher densities ($n_b\ge0.7$).

For cold NS matter,the npK EoS which was initially following np EoS, bends at the density point when antikaon condensates start appearing at $n_b\sim 0.5$.  Since there are no thermal kaons present at T= 0 MeV, it is only the presence of antikaon condensates which contributes to the softening of EoS.
For $s=1$  NS, the npK EoS follows the corresponding np EoS until thermal kaons enter, at which point the slope of the EoS curve changes slightly due to reduction in pressure. Further when the antikaon condensates appear at later density, the curve bends again further softening the EoS. The fraction of antikaon condensates becomes more than that of thermal kaons at about $n_b\sim 0.62$ and soon after we see EoS becoming so soft that T = 0 MeV EoS becomes stiffer than $s=1$  EoS. 

If we look at the expression for pressure in Eq. 5, contribution of the first three terms, i.e., the potential terms is more for a low $s$ NS matter. Even though with the rise in temperature, contribution of the kinetic terms increases, it dominates only at sufficiently high $s$ values ($s$ $\geq 2$) where thermal kaons play a significant role. 
In contrast, the contribution of potential term decreases with an increase in $s$ value. This decrease is maximum for low $s$ NS and minimum for high $s$ NS. This is because in a high $s$ NS, the thermal kaons appear at lower density and their fraction in the system is also relatively high compared to a NS at lower s; as will be seen in Figs. 4 and 5. On the other hand the fraction of $K^-$ is lower for a higher $s$ NS and they appear at higher densities only. 
In addition, the kinetic increase in pressure is also there due to high temperatures, so the overall EoS for high $s$ NS is stiffer than for a cold star. This, however, may become softer at higher densities where $K^T$ overtakes $K^-$ as was seen for the case of $s=1$ earlier. In general, we can say that a cold EoS which is initially softer than a higher $s$ EoS, may become stiffer at very high densities.

The EoS profile obeyed by a NS is reflected in its mass-density profile. We next study NS gravitational mass sequences for static as well as rotating stars obeying different EoS.
To get mass profiles of a static NS, we solve TOV equations for the same, with our sets of EoS. For the cold NS, the maximum masses of about 2.417 $M_{solar}$ and 2.372 $M_{solar}$ 
can be attained for np and npK matter respectively, after which the star becomes unstable as the slope, $\frac {d M_G}{d n_b}$, becomes negative. For hot, non-rotating stars,  Goussard et. al. (1997) \cite{Gouss}, based on the earlier papers \cite{Fried88, Sorkin82} have shown that a stable configuration can be distinguished from the unstable one following the gravitational mass sequences at constant total entropy $S = s *M_B$, instead of constant $s$. We use this criterion to find mass sequences for np and npK ($U_{\bar K}$=-100MeV) EoS
and plot gravitational mass-number density ($M_G-n_b$) profile for different entropy values in the right panel of Fig. \ref{fig1}. The plotted sequences are for $S$ of 0, 3, 5 and 7 $M_{solar}$ NS. 

The sequence for a $S$ = 3 $M_{solar}$ corresponds to NS configurations with an $s$ between 1 (for the higher $M_G$ end of the sequence) and 2 (for lower $M_G$ end). The maximum mass reached corresponds roughly to a NS at $s\sim 1.1$. Since the configurations at higher end are lower in $s$, we see very minor difference there between a cold NS sequence and a NS sequence at $S$ = 3 $M_{solar}$, essentially reflecting the EoS nature, where we saw hardly any difference between cold EoS and $s$ = 1 EoS. The higher $S$ sequences of 5 and 7 $M_{solar}$ correspond respectively to an $s$ of (1.8  - 3.8) and (2.5 - 5) respectively, where the values inside parentheses correspond to higher and lower end of the sequence.  These sequences are sharply different from the corresponding cold NS sequence, as the thermal effects are more dominant here. A high total entropy can sustain a higher gravitational mass as is evident from the figure.  For a given constant $S$, we get gravitational mass values until a limiting $s$ is reached, beyond which the sequence possibly enters the instability region. The maximum stable mass reached in a sequence as a result and the corresponding central number density for given $S$ of 0, 3 and 7 $M_{solar}$, are listed in Table ~\ref{tab1} for np and npK EoS with different $U_{\bar K}$. 
We observe that colder stars have a denser core than finite temperature NS. The central density further decreases as the entropy of a star increases. We see that, e.g., for a NS with $U_{\bar K}$ =-100 MeV, the mass increases from 2.297 $M_{solar}$ for cold NS to 2.346 $M_{solar}$ for a NS at $S=7M_{solar}$, whereas its central density decreases from 0.833 $fm^{-3}$ to 0.729 $fm^{-3}$. We refrain from quoting radius values for finite temperature stars as the surface pressure never really goes to zero.

\begin{table}
\caption{Maximum mass reached  for 
different np and npK EoS and the corresponding central number density at constant total entropy values of 0, 3 and 7 $M_{solar}$. The first row is for np EoS with no antikaon contribution and the rest five rows are for npK EoS at different antikaon optical potentials.}
\label{tab1}
\begin{tabular}{|c|c|cc|cc|cc|}
		\hline
        &$U_{\bar K}$&\multicolumn{2}{|c|}{S=$0M_{solar}$}&\multicolumn{2}{|c|}{S=$3M_{solar}$}&\multicolumn{2}{|c|}{S=$7M_{solar}$}\\
        &&$M_{max}$&$n_b$&$M_{max}$&$n_b$&$M_{max}$&$n_b$\\
        &MeV&($M_{solar}$)&($fm^{-3}$)&($M_{solar}$)&($fm^{-3}$)&($M_{solar}$)&($fm^{-3}$)\\
        \hline
        np&&2.417&0.851&2.416&0.829&2.437&0.740\\
        npK&-60&2.372&0.822&2.376&0.792&2.386&0.718\\
        &-80&2.339&0.823&2.341&0.799&2.369&0.727\\
        &-100&2.297&0.833&2.296&0.8&2.346&0.729\\
        &-120&2.242&0.862&2.237&0.825&2.315&0.731\\
        &-140&2.176&0.914&2.164&0.841&2.275&0.716\\
        &-150&2.142&0.95&2.125&0.912&2.251&0.722\\
		\hline
	\end{tabular}
   \end{table}
   
\begin{figure}
	\includegraphics[width=\columnwidth, height=8.5cm]{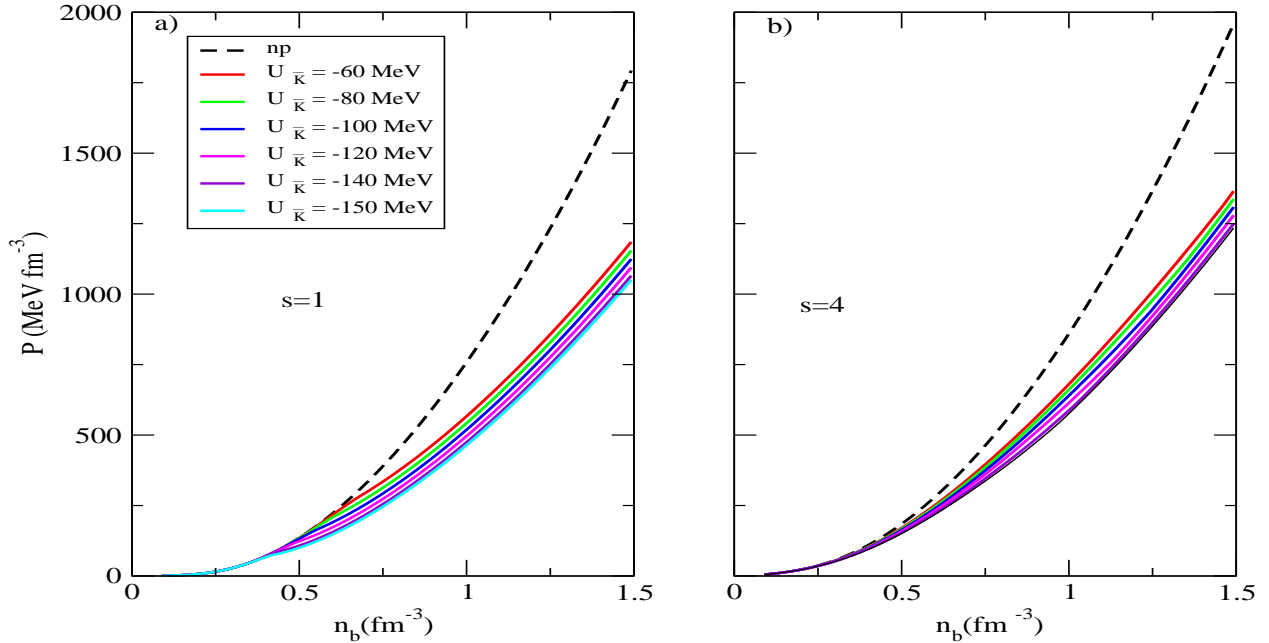}
    \caption{The EoS is plotted for a range of values of $U_{\bar K}= -60 MeV$ ~ to $-150 MeV$ a) for lower $s = 1$ and b) for higher $s = 4$. In both the plots np EoS is also included as the dashed line for comparison. }
    \label{fig2}
\end{figure}

\begin{table}
	\centering
	\caption{
The value of baryon density ($n_b$) in $fm^{-3}$ when $K^T$ and $K^-$ condensate start appearing in the NS core. They are for different values of $U_{\bar K}$ and also for different thermodynamic states inside the NS core.}

\label{tab2}

	\begin{tabular}{|l|c|cc|cc|cc|cc|cr|} 
		\hline
		$U_{\bar K}$ & T=0 &\multicolumn{2}{|c|}{s=1}  &\multicolumn{2}{|c|}{s=2 } & \multicolumn{2}{|c|}{s=3 } & \multicolumn{2}{|c|}{s=4 } & \multicolumn{2}{|c|}{s=5} \\
        MeV &  MeV & \multicolumn{2}{|c|}{ }&  \multicolumn{2}{|c|}{} & \multicolumn{2}{|c|}{ }& \multicolumn{2}{|c|}{} & \multicolumn{2}{|c|}{ }\\
		\hline
        &  $K^-$ & $K^T$ & $K^-$& $K^T$ & $K^-$& $K^T$ & $K^-$& $K^T$ & $K^-$& $K^T$ & $K^-$\\
        \hline
		-60 & 0.613 & 0.427 & 0.687 & 0.282 & 0.868 & 0.191 & 1.129& 0.131 & 1.396 &0.092& -\\
		-80  &  0.558 & 0.393 & 0.622 & 0.264 & 0.777 & 0.18 & 1.006 & 0.125& 1.245 & 0.089&-\\
		-100 &  0.507 &0.362 &  0.559 & 0.246 &  0.690 &0.171 & 0.889 & 0.121& 1.097 & 0.089&1.312\\
		-120  & 0.459 & 0.334 & 0.502 & 0.231&  0.608 &0.162 & 0.777 & 0.115 & 0.957 & 0.089&1.126\\
		-140  & 0.416 & 0.308  & 0.449 & 0.218  & 0.535& 0.155 & 0.674 &0.11& 0.828 & 0.089&0.969\\
		-150  & 0.395 & 0.297 & 0.425 & 0.21 & 0.449 & 0.15& 0.626 & 0.109&0.765 & 0.089&0.896\\
		\hline
	\end{tabular}
\end{table}


We next study the EoS profiles of hot NS having different $U_{\bar K}$ values. In Fig.~\ref{fig2}, we compare np and npK EoS profiles for two different values of $s$: $1$ and $4$.  The npK lines are plotted for a range of $U_{\bar K}$ from -60 MeV to -150 MeV for $K^-$ condensates in a nuclear medium. 
We notice that as soon as the thermal kaons enter the system, the slope of corresponding EoS changes. The EoS gets further softened at higher densities when the antikaon condensates appear. 
The densities at which the thermal kaons and the condensates appear in the NS core, for different $U_{\bar K}$ values, are listed in Table ~\ref{tab2}. For a given value of $U_{\bar K}$,  $K^-$ condensates appear at lower density for a cold NS whereas their appearance is delayed to higher densities for hot NS. Thermal kaons are not present in cold NS. For finite temperature NS, not only the $K^T$'s appear much earlier than $K^-$ but for higher $s$ they appear at lower densities.  The condensates do not appear in the system even at very high densities ($n_b\sim 1.3fm^{-3}$) for a high $s$ ($\sim 5$) NS, unless the optical potential is deep enough, $\abs {U_{\bar K}} \ge 100$ MeV. Hence, in a NS core, an increase in $s$ delays the $K^-$ onset in the system but advances the onset on $K^T$, thereby making the EoS stiffer. It is to be noted that np EoS is the stiffest among all considered here because neither $K^-$ nor $K^T$ exist in the system ever. In contrast, for a NS at given $s$, as the depth of optical potential increases, the condensates as well as thermal kaons start populating the core earlier, thereby making the corresponding EoS softer.

\begin{figure}
	\includegraphics[width=\columnwidth,height=8.5cm]{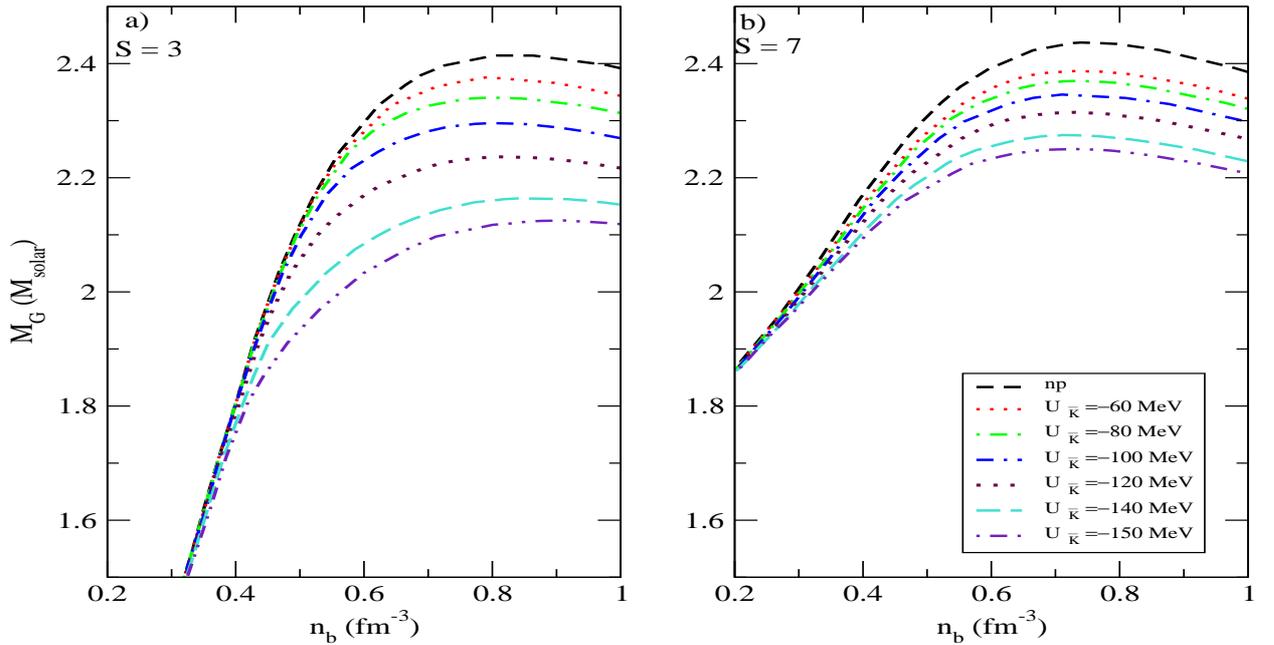}
    \caption{The mass-number density profiles for EoS with np and npK matter at different optical potentials at for a) S=3$M_{solar}$  and b) S = 7$M_{solar}$.}
    \label{figmn}
\end{figure}

Next, we study the NS mass sequences at constant $S$ for np and npK EoS.
The $M_G - n_B$ profiles are compared in Fig.~\ref{figmn} for constant total entropy of 3 $M_{solar}$ and 7 $M_{solar}$ in the two panels, which roughly translates to an $s$ of $\sim 1.1$ and $2.7$, respectively, for the higher $M_G$ end of a sequence.  
As before, nature of the EoS is reflected in the corresponding mass profile.
For the same thermodynamic condition, a softer EoS (deeper $U_{\bar K}$) makes a lower maximum mass star, 
The maximum masses reached by NS obeying a given EoS are listed in Table 1. This trend matches with earlier results \cite{pons, third}, however, the maximum mass values for cold NS are much higher and well above the observational constraint of $2 M_{solar}$ \cite{demo, anton, fons}.

\begin{figure}
	\includegraphics[width=\columnwidth, height=8cm]{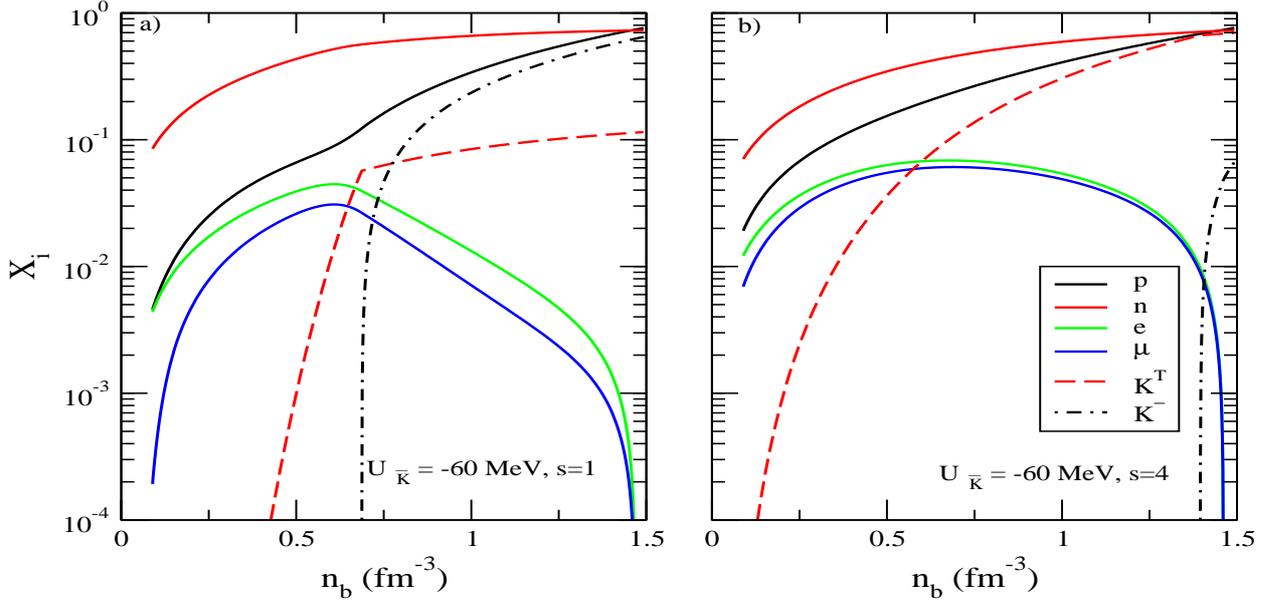}
    \caption{Fraction of different particles in a beta-equilibrated NS matter 
with  n, p, e, $\mu$  and antikaon condensates of $K^-$ and $K^T$; for 
$U_{\bar K} = -60$ MeV; plotted as a function of the baryon 
density for 
a) $s = 1$  and b) $s = 4$. }
    \label{fig2a}
\end{figure}

\begin{figure}
	\includegraphics[width=\columnwidth, height=8cm]{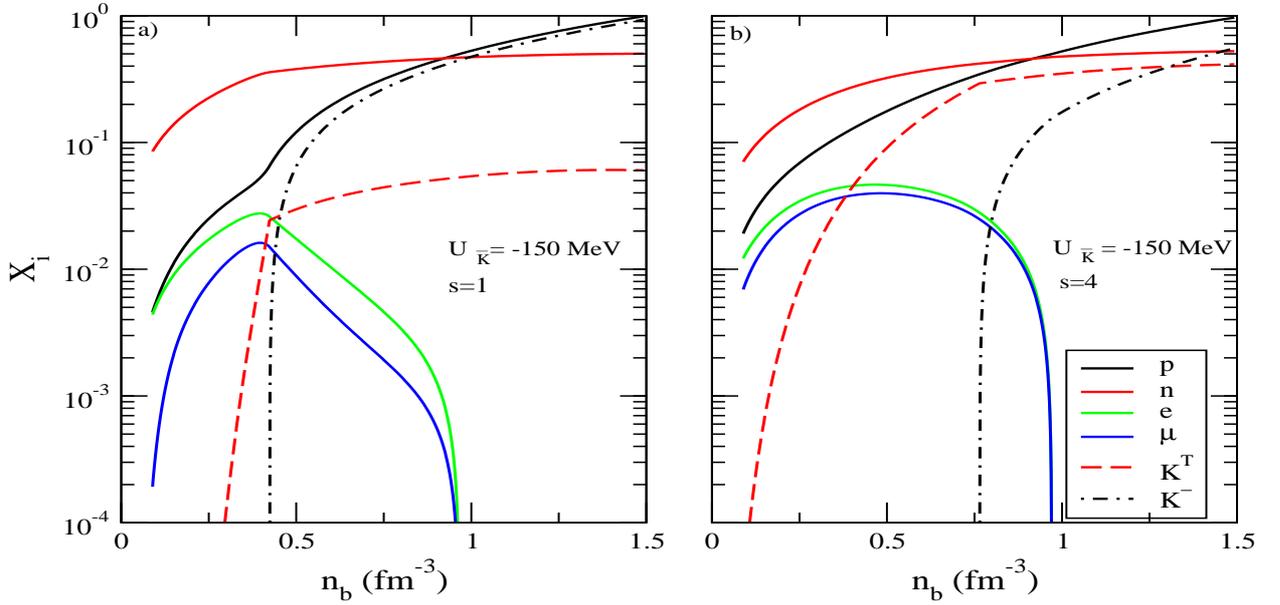}
    \caption{Particle fractions vs baryon number density as in Fig \ref{fig2a}, but for a deeper $U_{\bar K}= -150$ MeV for a) $s = 1$  and b) $s = 4$.}
    \label{fig2b}
\end{figure}

We next study the fraction of particles present in the NS core for EoS with different values of $U_{\bar K}$ and $s$. 
At low densities, the NS contains n, p, e and $\mu$. As higher densities are reached in a NS core, the threshold condition; $\mu_{K^-}= \mu_n- \mu_p= \mu_e$ is satisfied and $K^-$ appears in the system. The lepton fraction falls off as soon as the negatively charged condensates populate. The Bose-Einstein condensates do not contribute to the pressure and it is energetically favourable to have them in the system as compared to the leptons. We compare particle fractions in a hot NS with shallower $U_{\bar K} = -60 MeV$, for an $s$ value of $1$ and $4$ in the two panels of Fig. \ref{fig2a}.
Interestingly, in a low $s$ NS core, as soon as $K^-$ condensates appear, they quickly outnumber the already existing thermal kaons $K^T$. On the other hand, in a higher $s$ NS, the negatively charged $K^T$ appear at relatively lower density, pushing the onset of $K^-$ condensates to much higher $n_b$ values. $K^T$ was also noted to partially inhibit the appearance of the $K^-$ condensates in Pons et. al. (2000)\cite{pons}. In Fig. \ref{fig2b}, the particle fraction is plotted for a deeper $U_{\bar K}= -150 MeV$. A similar trend is noticed here as well. However, here the $K^-$ populates at lower densities compared to NS with shallower EoS. 
The threshold densities for onset of $K^T$ as well as for $K^-$ condensation  for different $s$ and $U_{\bar K}$ values were listed earlier in Table~2.

\begin{figure}
\begin{center}
\medskip
	\includegraphics[width=0.8\columnwidth, height=8cm]{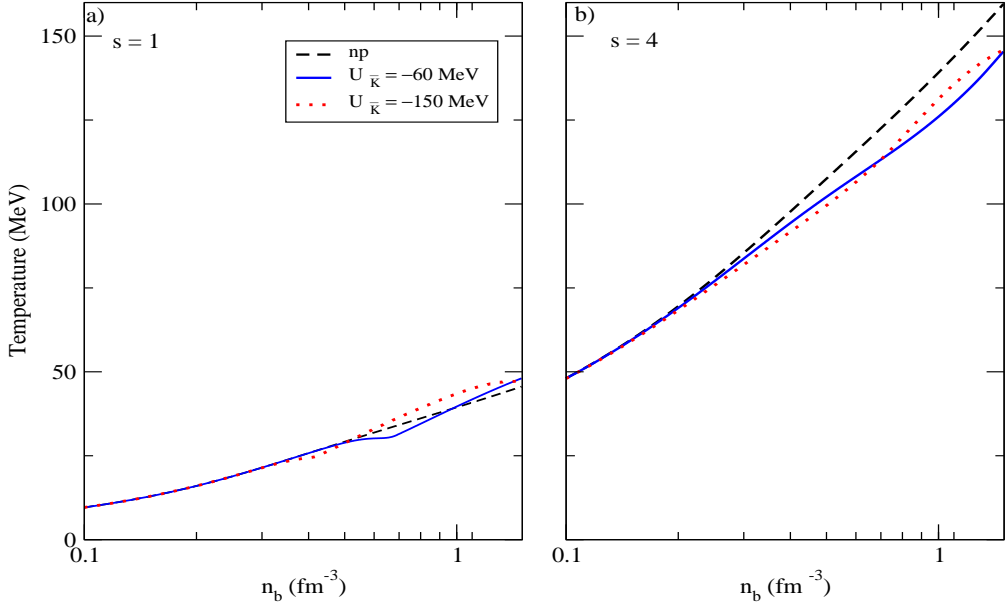}
    \caption{Temperature in a NS is plotted as a function of baryon density $n_b$, for a given thermodynamic state, a) $s = 1$ and b) $s = 4$.}
    \label{figT_n}
\end{center}
\end{figure}

Next we study the relation between temperature and number density (T vs. $n_b$) in a NS obeying different EoSs. We plot T vs. $n_b$ in the two panels of Fig. \ref{figT_n} for a NS with an $s$ of $1 $ and $4$ respectively. In each panel, we compare np with npK matter at shallow and deep optical potentials. 
For a fixed $s$, the temperature increases with increase in baryon density. In other words, the temperature falls off from the core of a NS to its surface. This nature of temperature curve was earlier reported in Banik et. al. (2008) \cite{PRC08} as well. At a given $s$, temperature is nearly the same at low densities for different EoS,  but in the high density region prominent difference can be seen between them which occurs because the thermal kaons and antikaon condensates start to populate the matter at these densities. We notice kinks in the npK lines which mark the appearance of $K^-$ condensates. For an isentropic NS with a high $s$ of $4$, the temperature can rise up to 150 MeV compared to 50 MeV for a star with low $s$ of $1$.  Also, in a lower $s$ NS, the core temperature is less for np matter as compared to that for npK matter.

For a lower $s$ NS, the antikaons not only appear at lower density, their fraction is higher for deeper potential EoS. Thus the core temperature rises for npK EoS with $U_{\bar K} = -150$ MeV, than for that of np EoS or npK EoS  at $U_{\bar K}= -60$ MeV. For npK at $U_{\bar K} = -60$ MeV and $s=1$, leptons depletion starts at the onset of thermal kaons (at $n_b = 0.427 fm^{-3}$) and the temperature curve of npK is lower than that for np EoS. The lepton depletion is accelerated  when condensates appear (at $n_b = 0.687 fm^{-3}$) and the temperature curve  surpasses that for np.
For a higher $s$ star the nature of temperature curve is  quite the opposite. Here, the core temperature of NS with np EoS is higher than that of npK. At higher s, the $K^-$ condensates appear at very high densities (see Table ~\ref{tab2}). Similar behavior is noted when additional fermionic degrees of freedom, such as hyperons are involved. In the absence of any variation of hyperon effective mass, it was shown that at a given baryon density a system with more components has lower temperature  \cite{proto}. 
%
\begin{figure}
\begin{center}
	\includegraphics[width=\columnwidth, height=8.5cm]{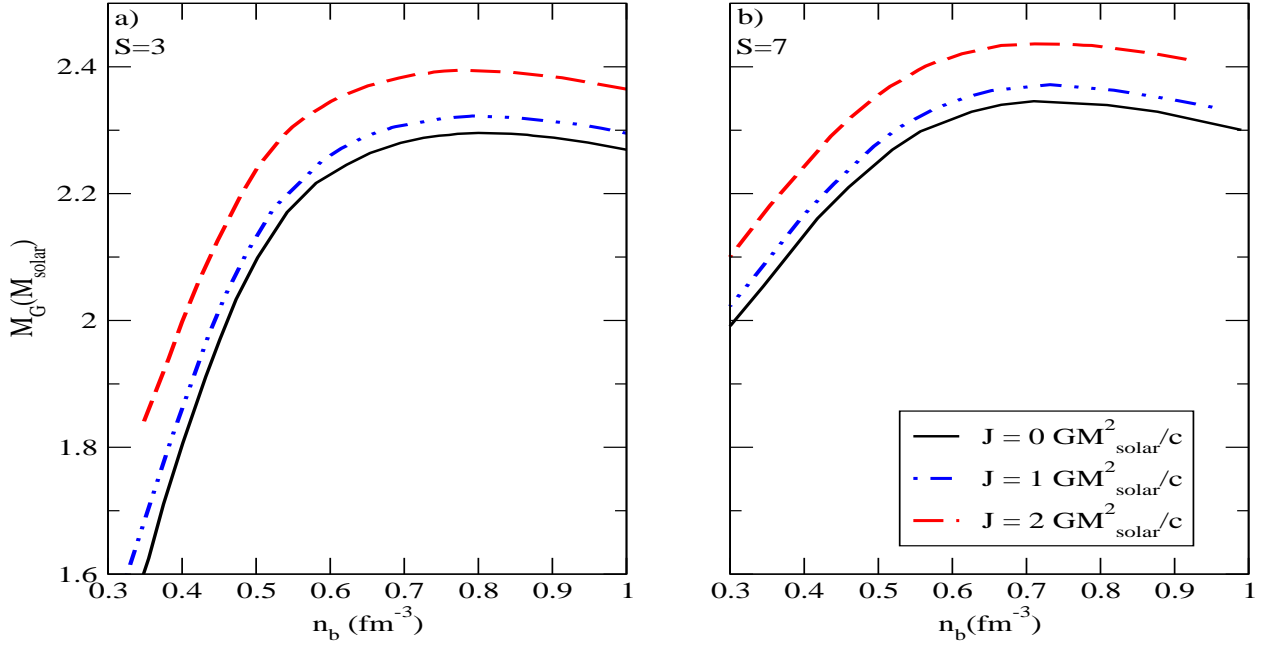}
    \caption{The evolution of mass with rotation for a NS with npK 
($U_{\bar K} = -100$ MeV) and at a) S = 3$M_{solar}$  and b) S = 7$M_{solar}$ . The 
sequences are 
plotted for NS rotating with different angular momenta starting from J=0 (static case) to $J=1 GM^2_{solar}/c$ and $2 GM^2_{solar}/c$. 
}
    \label{figmn_rot}
\end{center}
\end{figure}

Having discussed the properties of a static NS with np and npK EoS, we now study the rotating NS configurations using {\it Lorene/nrotstar}.  We certainly don't claim to give a completely realistic picture of a NS. This simplified picture of uniformly rotating isentropic, neutrino-less star at beta equilibrium is sufficient for the purpose of present work which is to study the influence of antikaon condensates on the properties of hot, rotating NS. A more complete study of neutrino-trapped PNS with our density-dependent EoS is left for future work. 
Keeping in consideration the observational pulsar frequency data,
we study the change in the NS configuration for different 
angular momentum values.

Fig.~\ref{figmn_rot} shows the evolution of gravitational mass$-$number density relation of a NS with change in its angular momentum. We plot the $M_G-n_b$ profiles of a NS for different angular momentum values $ (J= 0, 1$ and $2 GM^2_{solar}/c)$.
The mass sequences for EoS with npK for a moderate optical potential ($U_{\bar K}=-100 MeV$) plotted in the two panels are for a NS with an $S$ of $3$ and $7M_{solar}$, respectively.  
We see that a rotating star can support more mass compared to a static one. 
As the angular momentum changes from $0$ to $1 GM^2_{solar}/c$, the corresponding increase in $M_G$ is not very significant. But the relative change in $M_G$ for a NS with $J \sim 1 GM^2_{solar}/c$ to $2 GM^2_{solar}/c$ is significant, as is evident from both the panels of Fig. ~\ref{figmn_rot}.  The star with total entropy $7M_{solar}$ can support a maximum mass star which is more massive than that with total entropy $3M_{solar}$. The difference is independent of its angular momentum. However, we notice that the relative increase in NS mass from a lower J to higher J state is higher for a NS with lower $S$. The percent increase in NS mass was about 4.5 for S=3 NS sequence as we go from non rotating state to $J \sim 2 GM^2_{solar}/c$ state, whereas it was only 3.85 percent for S=7 NS for the same change in angular momentum state.

\begin{figure}
\hfill\begin{minipage}[t]{0.49\textwidth}
\includegraphics[width=11.2cm, height=8.4cm]{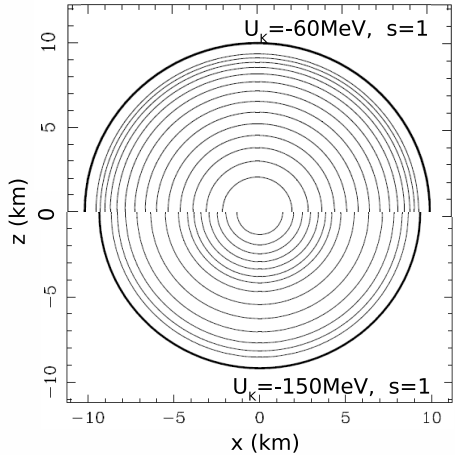}
    \caption{Energy density iso-contours of a 
rotating NS with baryon mass of $2M_{solar}$. 
Top(bottom) panel shows static NS with EoS for $U_{\bar K} = -60(-150)$ MeV, 
with $s = 1$.} 
    \label{contour1}
\end{minipage}%
\hfill\begin{minipage}[t]{0.49\textwidth}
\centering
    \includegraphics[width=10.8cm, height=8.0cm]{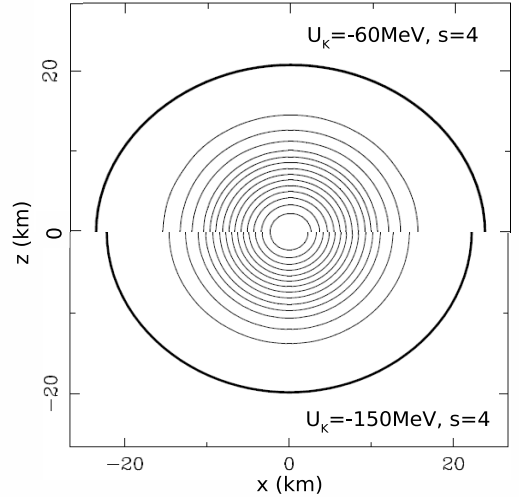}
    \caption{Energy density iso-contours of a rotating NS as in Fig \ref{contour1}, but with 
higher entropy per baryon state $s = 4$}.
    \label{contour2}
\end{minipage}
\end{figure}

We next study the change in shape of a NS for an increase in $s$ and for a change in its angular momentum. In Fig.~\ref{contour1} we compare a NS rotating at 300 Hz for two extreme npK EoS (with shallower $U_{\bar K}=-60 MeV$and with deeper $U_{\bar K} = -150 MeV$). The iso-contours lines drawn are of constant fluid energy density in the meridional plane, $\phi=0$. The vertical direction (y) is aligned with the stellar angular momentum. The thick solid line marks the stellar surface. The coordinates (x, z) are defined by $x = r sin \theta$ and $z = r cos \theta$, where $\theta$ is the polar angle. They represent the coordinate radii in x and z directions.
The upper slice of the Fig.~\ref{contour1} is for a NS with shallower EoS and bottom 
slice for EoS with deeper potential. Both NS are at a constant $s$ of $1$ and have same baryon mass of $2M_{solar}$.  As we can see from the figure, a deeper potential tends to make the NS more compact. Further, in both the cases, the NS are nearly spherically symmetric with the ratio of polar to equatorial radius $\sim$ 0.97.
Fig. \ref{contour2} gives contour plots for a NS with same EoSs (with shallower and deeper $U_{\bar K}$) as in the previous figure but with a higher $s$ of $4$. We notice that at higher $s$ the NS becomes bulkier, its size being almost double to that of a star with an $s$ of $1$. We also notice a slight deviation from the spherical shape for higher $s$ star with the ratio of polar to equatorial radius now $\sim$ 0.7. Thus, we conclude that the shape of a NS depends on its EoS and thermodynamic state. 
A NS with lower $s$ has a denser core and is more compact. Hence a higher $s$ star which is less compact deforms more when subjected to rotation as it gets bulged in the equatorial plane and is flattened in the vertical direction, which incidentally is also the direction of the stellar angular momentum.

\begin{figure}
\centering
\hfill\begin{minipage}[t]{0.47\textwidth}
\includegraphics[width=10.0cm, height=8.0cm]{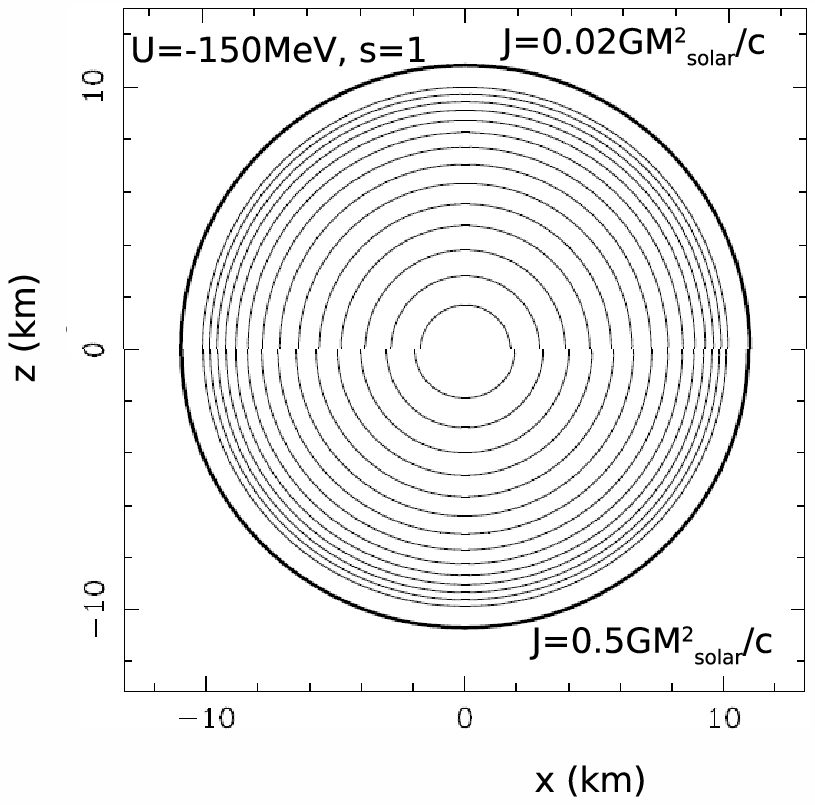}
\caption{Effect of rotation on NS shape. Energy density iso-contours 
for a NS with npK ($U_{\bar K}=-150$MeV, $s = 1$) rotating at 
J=$0.02GM^2_{solar}/c$  (top panel) and $0.5GM^2_{solar}/c$  
(lower panel).}
    \label{contour3}
\end{minipage}%
\hfill\begin{minipage}[t]{0.49\textwidth}
\centering
    \includegraphics[width=8.2cm, height=8.4cm]{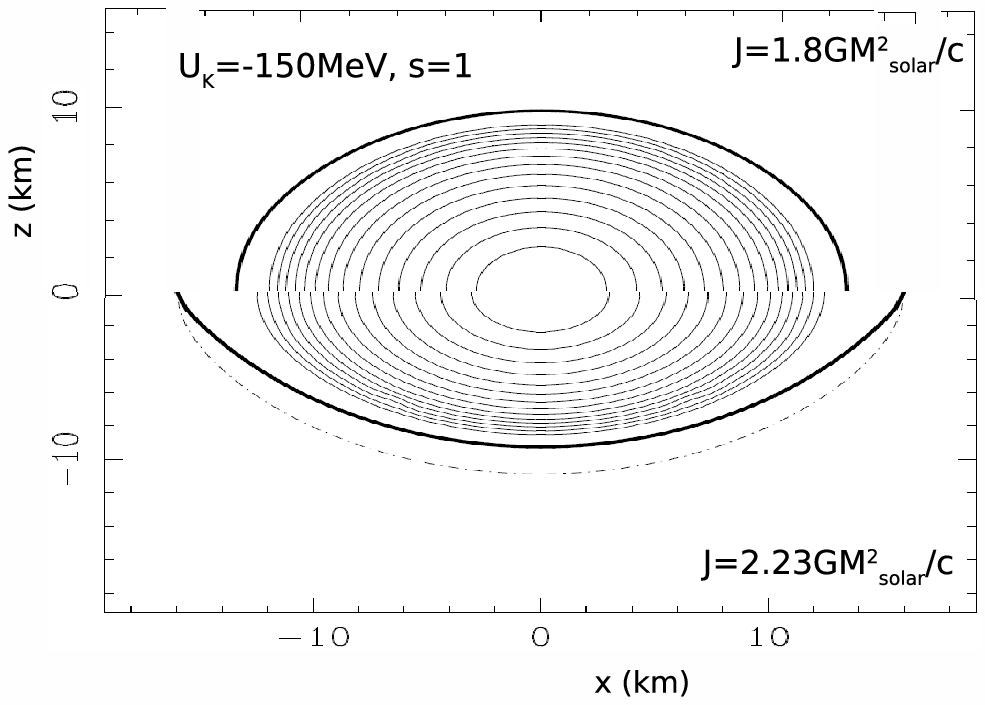}
    \caption{Energy density 
iso-contours for a NS as in Fig. \ref{contour3}
rotating at J=$1.8GM^2_{solar}/c$  (top panel) and $2.23GM^2_{solar}/c$  
(lower panel).}
    \label{contour4}
\end{minipage}
\end{figure}

\begin{figure}
\centering
\includegraphics[width=9cm, height=7cm]{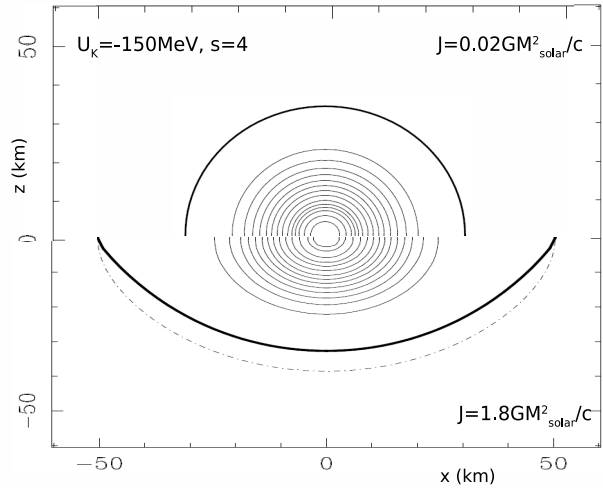}
\caption {Fluid energy density iso-contours for NS ($U_{\bar K}= -150$ MeV, $s = 4$) 
rotating at J=$0.02 GM^2_{solar}/c$ (top panel) and 
at J=$1.8GM^2_{solar}/c$ (lower panel)  }
    \label{contour5}
\end{figure}

Next we study the effect of rotation on a particular NS 
configuration. Top slice of Fig.~\ref{contour3} shows a NS with npK matter at deeper potential and low $s$ ($U_{\bar K}= -150$ MeV, $s = 1$. The NS is rotating slowly at 11 $Hz$ and has an angular momentum J$\sim 0.02$ $GM^2_{solar}/c$. NS shown in bottom slice rotates slightly faster at 280 Hz and has J$\sim0.5$ $GM^2_{solar}/c$. Fig. \ref{contour4} has the same NS with J$\sim$1.8 (top slice) and $2.23$ $M^2_{solar}/c$ 
 (bottom slice) and rotating with a frequency of 830 Hz and 930 Hz (Keplerian) 
respectively.  The NS is fairly spherical at low angular momenta, 
but gets deformed at higher $J$. At Keplerian frequency, the NS becomes elongated in an effort to keep itself from falling apart.
Thus the rigid rotation of a NS changes its shape as well as its equatorial radius. 
In the final contour plot of Fig.~\ref{contour5} we check the 
deformation of the star with npK EoS at higher $s$ of $4$.  We compare a slowly rotating  star having $J= 0.02GM^2_{solar}/c$ with a fast rotating star having $J= 1.8 GM^2_{solar}/c$. We observe the same pattern of deformation in shape at higher angular 
momentum as was noted for Figs. ~\ref{contour3} and \ref{contour4}.  
However, the deviation from spherical symmetry is much more 
pronounced in this case. This can again be attributed to the 
lower density of the core for a higher $s$ NS. 

In a NS, the deviation from spherical symmetry due to the anisotropy of energy-momentum tensor in the presence of strong magnetic fields has been reported by several authors. It has also been shown that the inclusion of magnetic field effects in the EoS and the interaction between the magnetic field and matter (or magnetisation) does not affect the stellar structure considerably \cite{Chat15,Fran16}. Without considering the magnetic field effects in our EoS, we made an order estimation for the GW emitted due to non-axisymmetric deformation or the ellipticity in a magnetized NS with baryon mass 2$M_{solar}$. 

We consider a typical NS with dipolar magnetic field that is uniform inside the star. For an estimation of GW amplitude, we assume that the magnetic and rotational axes are not aligned. Also, the star is assumed to be rotating at a frequency of 200 Hz which is much lower than its mass shedding limit, such that the deformation is primarily due to the strong magnetic field. Further, we assume that the magnetic energy is much less than the rotational kinetic energy as it is considered to be a realistic case. The NS then becomes a triaxial ellipsoid and emits GW as a result \cite{Bona96}. With this, a numerical estimate of the GW amplitude can be made using the following relation \cite{Bona96},

\begin{equation}\label{eq_h0}
h_0= 4.21 \times 10^{-24} \Big [\frac {ms} {P}\Big]^ 2 \Big[\frac {kpc} {D} \Big] \Big[\frac {I} {10^{38} kg m^2}\Big] \Big[\frac \epsilon {10^{-6}}\Big].
\end{equation}

Here, P is the rotation period of the NS, {\it D} is the distance to the NS, {\it I} is its moment of inertia with respect to its rotation axis and $\epsilon$ characterises the ellipticity of the NS due to magnetic field induced distortion. For a NS with polar magnetic field $B_{pole}$, the ellipticity is given by \cite{Bona96}

\begin{equation}\label{eq_ell}
\epsilon = \frac {45} {64\pi} \frac {B_{pole}^2} {\mu_0 G \rho^2 R^2}.
\end{equation} 
To make an order of magnitude estimate, we consider a typical magnetar formed in a binary merger event that is at a distance of $D = 40 Mpc$\cite{GW170817} . We assume that the NS has a polar magnetic field, $B=10^{15}$ G.  
Using Eqs. \ref{eq_h0} and  \ref{eq_ell} we then obtain a GW amplitude 
$h_0 \sim 9.3 \times 10^{-30}$ for a cold NS (T= 0 MeV). It increases marginally to 
$1.09 \times 10^{-29}$ for hot NS with $s = 1$ and $4.18 \times 10^{-27}$ for NS with $s = 4$.  Since all the above calculations assume the same B, the difference in GW amplitude is due to the thermal effect on EoS only. The GW emission from such a NS occurs at frequencies of $f$ and $2f$ (where $f$ is the rotation frequency of the NS). The strength of the two components is determined by the angle $\alpha$ between the distortion axis (axis of magnetic field induced distortion) and the rotation axis of the NS. For small $\alpha$, $f$ is the dominant component and for large $\alpha$, $2f$ component dominates \cite{Bona96}. For our case, the GWs from the distorted NS will be emitted at 200 Hz and 400 Hz, with the relative strength of each component being determined by the angle $\alpha$.

From the sensitivity curve of the present day detectors \cite{GWplotter}, it 
can be seen that the possibility of detecting a GW of this amplitude is 
severely limited. VIRGO and aLIGO, for example, have the range from about 10 Hz to a 
few kHz and the characteristic strain window of $\sim 10^{-22}$ to few times $10^{-24}$.
The next generation of ground based interferometers such as the Einstein Telescope are 
predicted to have a sensitivity that will bring down the characteristic strain 
down to about a few times $10^{-25}$ \cite{GWplotter}. Our calculations assume a 
rigidly rotating NS that is made of incompressible fluid and that has a uniform 
magnetic field inside the star. Relaxing these assumptions may lead to a 
greater value of ellipticity and hence a higher GW amplitude, 
which will have a high probability of being detected by the future generation 
of GW interferometers. 

On the other hand, many studies of CCSN simulations of PNS evolution have shown 
that during the early phase of PNS evolution after core collapse, the GWs are emitted via quasi normal modes and are expected to have frequencies of about a kHz and amplitude 
that lies well within the range of ground based detectors  \cite{Shibata04, Ferrari03,Moro18,Came17}. Also, many multidimensional CCSN simulations \cite{Fog15,Janka16} have shown g-modes as an important imprint of PNS oscillations in the early stages after bounce. The GW emitted as a result are expected to be about a few hundred Hz in frequency \cite{Sotani} having amplitude that should be in the grasp of current and upcoming GW observatories. However, CCSN explosion is a rare event, for example, in our galaxy  it happens  at the rate of about two to three times per 
century. It has also been shown by numerical simulations that short-lived but supra-massive neutron stars can be formed by the coalescence of low compactness NS of nearly equal mass \cite{Shibata17}. These rapidly rotating and highly non-axisymmetric products of merger are supported by differential rotation and would emit quasi-periodic 
GW with typical frequencies of about 2-3 kHz with substantially higher amplitude\cite{Bauswein17, Rezzolla17}.  Simulations have shown that up to $0.01 M_{solar}$ could be radiated in GW via this mechanism \cite{Ferrari03}. We intend to study these scenarios in future.

\section{Conclusions}
\label{sec:concl}
In the present paper we studied the set of NS EoS that contain  thermal kaons ($K^T$) and $K^-$ condensates in its core. This is done within the framework of relativistic mean-field theoretical model with density dependent couplings. We also compare these EoS with nucleon-only EoS. All of these have been studied for a set of constant $s$ NS. The finite $s$ NS is then compared with NS at zero temperature. 

The EoS with exotic matter tends to be softer as compared to np EoS. Moreover, among the npK EoS with antikaon condensates at different $U_{\bar K}$, the EoS with deeper potential makes $K^-$ condensates appear at lower densities in the core than that for a shallower $U_{\bar K}$, thereby resulting in softer EoS for matter at deeper optical potential. In general, the npK EoS also appears to stiffen as the entropy per baryon of a NS core increases.  Our static results are qualitatively consistent with earlier work of Pons et. al. (2000), where the EoS of kaon-condensed matter including the effects of temperature and trapped neutrinos were throughly studied \cite{pons}. 
The set of cold EoS we studied, however, fall within the required observational limit of 2 $M_{solar}$ star unlike theirs. 
 
We also studied the fraction of various particles in a NS core obeying a given EoS and noticed that the fraction of thermal kaons increases with an increase in $s$  but decreases slightly with an increase in potential depth, whereas the fraction of antikaon condensates decreases with $s$ but increases with an increase in the depth of antikaon optical potential.
 
We next studied the mass sequences for NS at constant total entropy $S$ and found that the EoS behaviour is closely reflected in these mass sequences. It was observed that maximum mass of a NS sequence increases with an increase in $S$. In contrast, the maximum mass attained in a sequence decreases as the depth of optical potential increases. We next studied the evolution of the mass-number density relation with various angular momenta for different EoS. The maximum mass of a given sequence was found to increase with an increase in the corresponding angular momentum.

We also observed the fall of temperature from the core to the surface of a NS. The core and the surface temperature depend on the EoS and the thermodynamic state of matter.or low $s$, the temperature in NS rises in the presence of antikaons. However, the temperature is more for np matter compared to npK matter for higher $s$ NS. In the presence of other exotic fermions, such as hyperons, this trend was reported in earlier work \cite{proto} as well.

Further, we studied the effect of rotation on the equilibrium structure of a NS in the form of iso-contours of its fluid energy density. A NS obeying an exotic EoS with deeper  potential tends to be more compact as compared to a NS with an EoS with shallow $U_{\bar K}$. In both the cases, we find that at low $J$, a NS tends to be nearly spherically symmetric but starts to deviate from spherical symmetry as its $J$ increases. The NS deforms considerably as its reaches the Keplerian limit. The deformation was found to be  more for a higher $s$ NS. Also, a higher $s$  star has less dense core and thus bulges more in the equatorial plane when subjected to fast rotation along the polar axis.

Finally, we made a crude estimate of the GW amplitude for a highly magnetised NS whose magnetic axis is not aligned with its rotation axis. The GW amplitude for a hot NS with high $s$ was found to be considerably larger than that for a cold NS. Still its strength is not large enough to come in the range of the present day detectors, but might just come within the grasp of next generation of GW observatories.

Results of differentially rotating configurations of NS and the limits of GW emission as a result of the instabilities triggered by mass-shedding limit as well as strong magnetic fields  will be reported subsequently.

\section*{Acknowledgements}
The authors gratefully acknowledge Department of Science \& Technology, 
Government of India for the financial support. NDB is supported by DST Grant no. SR/WOS-A/PM-1031/2014(G). 
Research of KPN and SB is supported by DST Grant no. SB/FTP/PS-205/2013. Also, the authors would like to thank to M. Oertel, S.S. Lenka and P. Char for useful discussions.
Finally they thank the anonymous referee as his/her comments have helped in 
understanding as well as improving the paper to a large extent.
\newpage

\end{document}